# Length-scale study in deep learning prediction for non-small cell lung cancer brain metastasis

Haowen Zhou[1,†], Steven (Siyu) Lin[1,†], Mark Watson[2], Cory T. Bernadt[2], Oumeng Zhang[1], Ramaswamy Govindan[3], Richard J. Cote[2,*], Changhuei Yang[1,*]

[1] California Institute of Technology, Department of Electrical Engineering, Pasadena, CA. 91125, USA

[2] School of Washington University Medicine, Department of Pathology and Immunology, St. Louis, MO. 63110, USA

[3] Washington University School of Medicine, Department of Medicine, St. Louis, MO. 63110, USA

[†] The authors contribute equally to this work

[*] Corresponding authors: chyang@caltech.edu, rcote@wustl.edu

**Abstract**

Deep learning assisted digital pathology has the potential to impact clinical practice in significant ways. In recent studies, deep neural network (DNN) enabled analysis outperforms human pathologists. Increasing sizes and complexity of the DNN architecture generally improves performance at the cost of DNN's explainability. For pathology, this lack of DNN explainability is particularly problematic as it hinders the broader clinical interpretation of the pathology features that may provide physiological disease insights. To better assess the features that DNN uses in developing predictive algorithms to interpret digital microscopic images, we sought to understand the role of resolution and tissue scale and here describe a novel method for studying the predictive feature length-scale that underpins a DNN's predictive power. We applied the method to study a DNN's predictive capability in the case example of brain metastasis prediction from early-stage non-small-cell lung cancer biopsy slides. The study highlights the DNN attention in the brain metastasis prediction targeting both cellular scale (resolution) and tissue scale features on H&E-stained histological whole slide images. At the cellular scale, we see that DNN's predictive power is progressively increased at higher resolution (i.e., lower resolvable feature length) and is largely lost when the resolvable feature length is >5 microns. In addition, DNN uses more macro-scale features (maximal feature length) associated with tissue organization/architecture and is optimized when assessing visual fields >41 microns. This study for the first time demonstrates the length-scale requirements necessary for optimal DNN learning on digital whole slide images.

# 1 Introduction

Digital pathology with deep learning analysis is of increasing importance in pathologic research and clinical practice [1], [2]. New developments in imaging technology and artificial intelligence (AI) hold the potential to ease certain laborious tasks in clinical diagnosis [3], [4]. More intriguingly, deep learning enabled digital pathology analysis has demonstrated the capability to make clinically relevant diagnoses based on subtle image features that can outperform human pathologists and that defy human interpretation. The performance of such deep neural network (DNN) based analysis systems generally scales with its size and complexity which further exacerbates the ability to understand the broader clinical interpretation of AI-identified pathology features that may provide physiological disease insights.

To improve interpretability, numerous novel approaches have been developed and evaluated in recent years [5]–[8]. Saliency maps or class activation maps can be computed to visualize the attention of DNN from backpropagating the model weights [9]–[12]. The intuition behind this type of method is that the change in the DNN output can be traced back to the individual perturbations of the pixel values in the images. These methods can highlight where the DNN is focusing and give humans a sense of which part might be more important within the image. A second group of saliency mapping methods is based on occluding some parts of the images and checking the resulting variation in DNN outputs [13], [14]. Similarly, occlusion maps can be generated to track DNN's attention. The third category of methods is to visualize the digital filters in different layers of DNNs [15]–[17]. In medical image analysis, other explainable models have been developed such as concept learning models [18]–[20] and case-based models [21], [22]. The concept learning models predict high-level clinical features first, and then make final decisions based on these clinical features. Case-based models make predictions by comparing the latent space features extracted from an input image against class discriminative prototypes. The methods mentioned above have achieved significant advances in explaining DNN for human-understandable tasks, such as image classifications [12], [23], tissue or cell segmentation [24], and auto-pilot of self-driving systems [25].

In this paper, we introduce a new method for shedding light on the prediction mechanism in image-based DNN analysis. This approach operates by altering the image resolution and field-of-view of the training and testing data set and studying the impact on DNN prediction accuracy to identify the length-scale that optimize DNN learning. This generalized approach is applicable to all DNN architecture, as it is network agnostic. In the application area of digital pathology analysis, it provides length-scale information about the DNN's predictive ability which can be used to optimize that ability and that a human pathologist can in turn use to better understand predictive physiological features.

Here, we demonstrate the use of this method on a DNN that has been trained to predict brain metastasis in early-stage non-small-cell lung cancer (NSCLC) subjects through high-quality whole slide images (WSI) of the diagnostic pathology slides. For context, we note that NSCLC is one of the most lethal cancers globally. Nearly a third of early-stage (Stage I-III) cases will recur with distant metastases [26], and brain metastasis is a common cause of morbidity and mortality in NSCLC [27]. At present, it is not possible to accurately predict the metastatic potential of NSCLC using conventional histopathological analysis, even when supplemented with genomic or molecular biomarkers [28]. This limitation is especially significant for patients diagnosed at an early stage, where precise risk assessment plays a pivotal role in making treatment decisions. Recently, we developed and trained a DNN to predict brain metastasis using diagnostic WSI [29]. While this DNN is capable of making meaningful and statistically significant predictions, it shares the opacity associated with most other related image analysis DNN – thus making it an excellent candidate for this length-scale study method.

The length-scale study is composed of training and testing the same DNN for different resolvable feature lengths (RFLs) and maximum feature lengths (MFLs) of the input images. The detailed definition of RFL and MFL will be introduced in the Material and Methods section. The RFLs reveal feature detail length-scale that is important to the DNN predictive ability, while the MFLs reveal how distance and large-scale features contribute to the DNN outcome.

## 2 Material and Methods

This study is based our recent demonstration that DNN on WSI from patients with early stage NSCLC can reliably predict which cases progressed to brain metastases and those that had no recurrence of any type after extended follow-up [29]. The current study uses the same WSI to investigate features necessary to optimize the prediction potential of DNN.

2.1 Patient Cohort

A cohort of treatment-naive 158 patient cases were recruited in this study with early-stage (Stage I-III) NSCLC diagnosed and treated at Washington University School of Medicine with long-term follow-up (>5 years or until metastasis). In total, 158 fresh Hematoxylin and Eosin (H&E) stained slides were retrieved and processed from existing, de-identified formalin-fixed paraffin-embedded (FFPE) diagnostic tissue blocks and imaged by an Aperio / Leica AT2 slide scanner. The 158 cases were categorized into two groups, Met+: 65 cases with known brain metastasis, and Met-: 93 cases with no recurrence after extended follow-up. The median follow-up time of this cohort was 12.2 months and 106 months for Met+ and Met- groups, respectively.

2.2 General deep learning pipeline

The scanned 158 whole slide images were first reviewed by a pathologist. The primary tumor regions with their tumor micro-environment were roughly annotated (Fig. 1(a)). The Ostu thresholding [30] was performed at the annotated regions to remove the backgrounds (plain glass). A thousand nonoverlapping image tiles for every whole slide image were randomly selected to undergo a color normalization process [31], which are then taken as the inputs for the DNN.

The ResNet-18 [32] convolutional neural network pretrained on the ImageNet dataset was used as the backbone in our DNN structure. The linear layers in the DNN were replaced by a linear layer and a sigmoid activation layer to adapt our binary classification task. For every input image tile, the model outputs an individual score – a prediction score. This tile-level score is supervised by the binary-encoded label from Met+ and Met-.

In clinical deep learning studies, due to a lack of well-established testing sets, multiple training-testing splits are usually adopted to avoid potential bias in the testing set selection from a single experiment. As illustrated in Fig. 1(b), we designed three individual experiments with different training-testing splits. Each of the training sets had 118 cases (Met+: n=45, Met-: n=73) and each of the testing sets had 40 cases (Met+: n=20, Met-: n=20), with 1000 image tiles per case. Note that the mild imbalanced training dataset effect was mitigated by using a large batch size with 200 image tiles in each batch [33]. Therefore, in each gradient descent iteration, the model saw adequate samples in both Met+ and Met-. Dataset imbalance is a common situation in machine learning, and the protocols for addressing it are common and numerous [33]–[38]. The training-testing splits were done with randomization and the three testing sets had no overlapping patients. Finally, the performance of the classifier is evaluated by the tile-level accuracies.

2.3  Length-scale definition and investigation

The RFL is the minimum length that can be resolved in the input image. This feature length is constrained by the Nyquist-Shannon sampling theorem [39] in the digital image. Here, we take a full resolution input image tile as an example. The input image has a pixel count of 224-by-224 pixels with a pixel pitch of 0.51 microns. According to the Nyquist-Shannon sampling theorem, the minimum resolvable feature required at least two pixels in image sampling along each of the lateral axes. In this case, our RFL is 1.2 microns which is twice of the pixel pitch. In this example, the RFL matches the physical resolution limit of the microscope used to acquire this image. By down-sampling the image while maintaining the same tile size, we can generate images that are more poorly resolved and thus have larger RFL. RFL can alternately be interpreted as the image resolution – we avoided using the term 'resolution' here because resolution generally refers to one of the imaging system specifications, while we use RFL here to refer to the target sample intrinsic feature length-scales.

Our study involves performing a sequence of down-sampling to the base image data set in order to generate data sets with varying RFL. As a technical note, we note that to avoid passing tiles of different pixel counts to the DNN and complicating the comparison process, the image tiles are interpolated after down-sampling to maintain the same overall pixel counts.

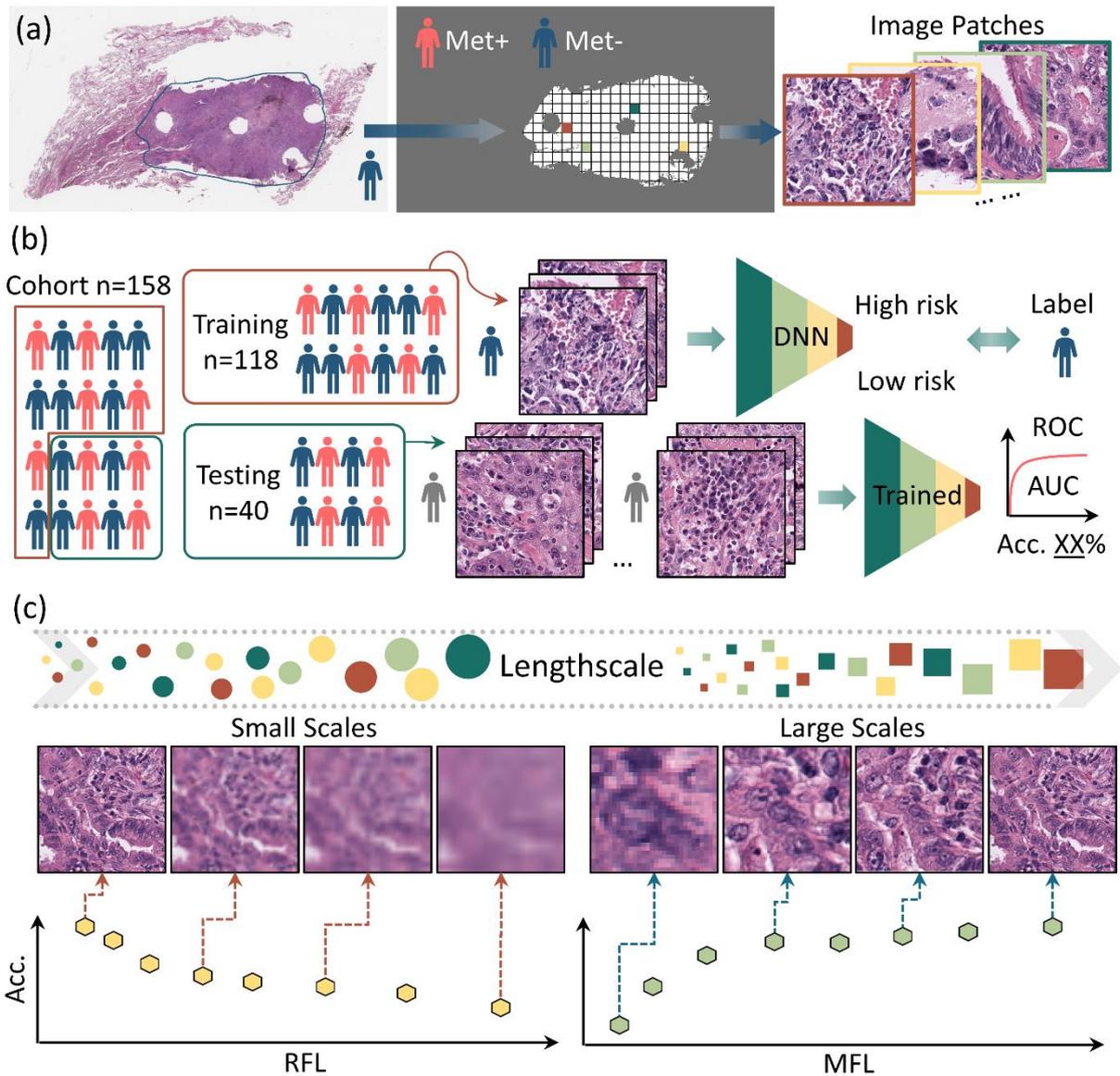

Fig. 1 (a) Preprocessing pipeline of H&E-stained whole slide images. The whole slide image is manually annotated by a human expert. The annotation mask is processed with thresholding to get rid of the background region. A thousand non-overlapping image tiles are randomly selected from the masked region. (b) One experiment of training-testing split in deep learning pipeline. Acc. stands for "accuracy". (c) An illustration of the length-scale study. The sizes of the circles and squares indicate the sizes of the feature scales.

The maximum feature length (MFL) is associated with the input image field-of-view. Specifically, the MFL is confined by the images' physical support – the size of the image in microns or the lateral extent of the image field-of-view. Once again, we can use the input image with a pixel count of 224- by-224 pixels and a pixel pitch of 0.51 microns as an example. In this case, the MFL is 114 microns from the product of 224 pixels and 0.51 microns per pixel. We can generate data sets with smaller MFLs by using smaller tile sizes. As a technical note, we again note that to avoid passing tiles of different pixel counts to the DNN and complicating the comparison process, the image tiles are interpolated after down-sampling to maintain the same overall pixel counts.

The length-scale study was conducted under different RFLs and distinctive MFLs for the input image tiles (Fig. 1(c)). For RFLs, the image tiles were resampled with fewer pixels in rows and columns and then interpolated back to the original size. After operating this process, the detailed features of the image contents were removed. For different MFLs, the image tiles were cropped to smaller sizes and then interpolated back to the original size to meet the input size requirement from DNN. In this process, the distant features or large-scale features were blocked out before feeding into the DNN. The combination of these two processing can reveal contributions from various levels of the features. For every RFL or MFL, the deep learning pipeline described in Section II-B was performed. The results and interpretations will be discussed in the Results section.

## 3  Results

The initial pilot study was designed to check the feasibility of DNN in predicting brain metastasis [29]. Though brain metastasis occurs frequently in NSCLC, no clinically effective predictor has been reported, particularly in early-stage (Stage I-III) NSCLC [28], [40]. The pilot study DNN was trained on a relatively homogenous patient cohort with well-defined and clinically relevant endpoints (Met+ versus Met-). The DNN was evaluated/tested on three training-testing experiments.

We preserved the architecture of the DNN from the original pilot study. Our length-scale study consists of a sub-study on the RFL and a sub-study on the MFL. For the RFL sub-study, we down-sampled the input image tiles from 1 to 30 in 18 levels compared to their original size, which corresponds to the RFL range of 1.2 microns (the sharpest resolution we could obtain for our images based on image acquisition at 20×) to 31 microns. Image tiles at each RFL value underwent the threefold training-testing procedure described in Section II-B. The average achieved accuracy for the experiments versus RFL is plotted in Fig. 2(a). For the MFL sub-study, we prepared 12 data sets with MFL ranging from 2.5 to 114 microns. The image tiles at each MFL value underwent the same threefold training-testing procedure described in Section II-B. The average achieved accuracy for the experiments versus MFL is plotted in Fig. 2(b).

We note that the leftmost points in Fig. 2(a) are the same as the rightmost points in Fig. 2(b). They represent the prediction accuracy of the DNN on unmodified tiles in the original studies. As we degrade the information content of the tiles (either by increasing RFL or decreasing MFL), we see decreases in prediction accuracies in the figures due to the different mechanisms of information loss (either from RFL or MFL).

The average values of the three experiments in Fig. 2 were fitted with piecewise linear functions with different slopes and intercepts, which exhibits a smaller fitting error compared to regular linear fitting. The variations in tile-level accuracy changes across RFLs and MFLs indicated that distinctive length-scales contribute differently to the overall accuracies. We noticed that when we significantly increased RFL or reduced MFL, the accuracy did not drop to 50% as a random guess in binary classification. This suggests that the color or staining of these H&E images likely have a slight impact on the predictivity of brain metastasis.

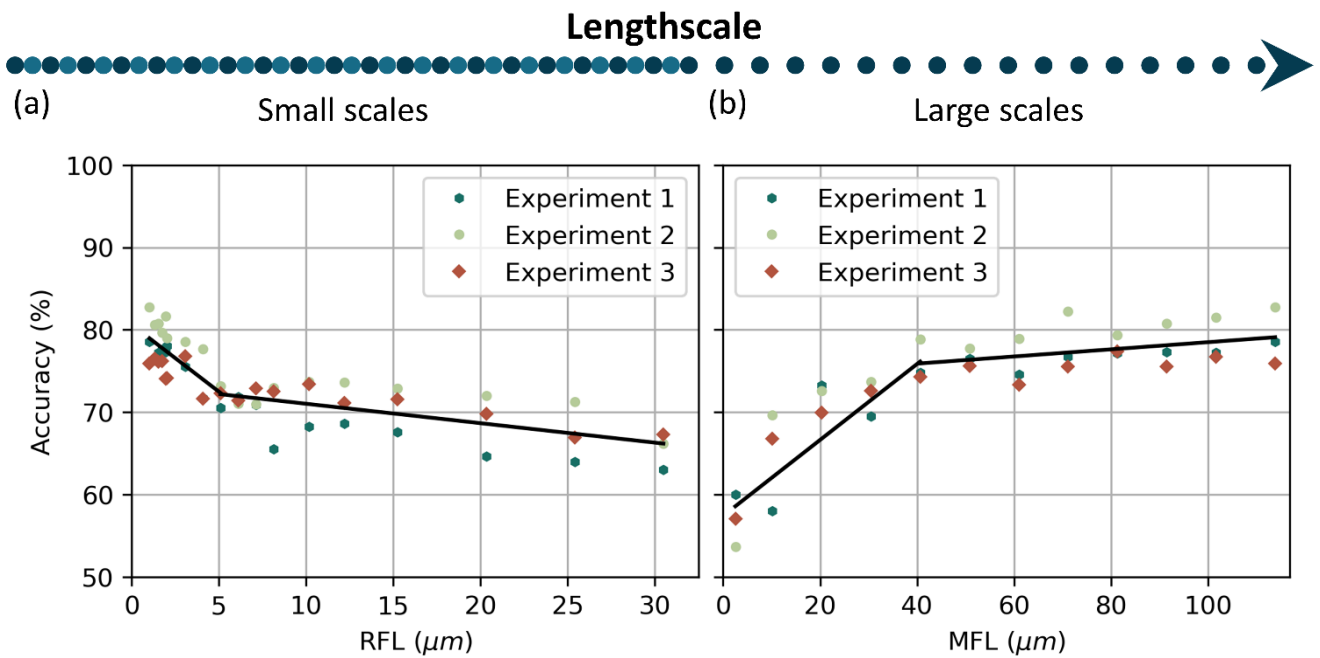

Fig. 2 Length-scale study curves for different (a) RFLs and (b) MFLs. The black solid lines are the piecewise linear fittings to the average values of the three experiments.

The curves for accuracy versus RFL underwent a relative rapid drop before leveling out. The transition point was determined to be at a RFL of 5.1 microns – for ease of reference, we will label this as the characteristic RFL. This transition point was determined by minimizing the absolute residue error from the piecewise linear regression. The sharp drop-off in prediction accuracy before the characteristic RFL suggests that the DNN's predictive capability is most sensitively associated with features as length-scale smaller than 5.1 microns. As can be seen in Fig 3(a), the accuracy markedly improves as length scale feature become progressively smaller (i.e., progressively achieve higher visual resolution (Fig 1c) between 5.1 microns and 1.2 microns (maximal resolution). That is, the higher the resolution of the image, the greater the prediction accuracy. This result also suggests that if we could have obtained smaller RFL, by for example scanning at higher power, even greater accuracy could have been achieved.

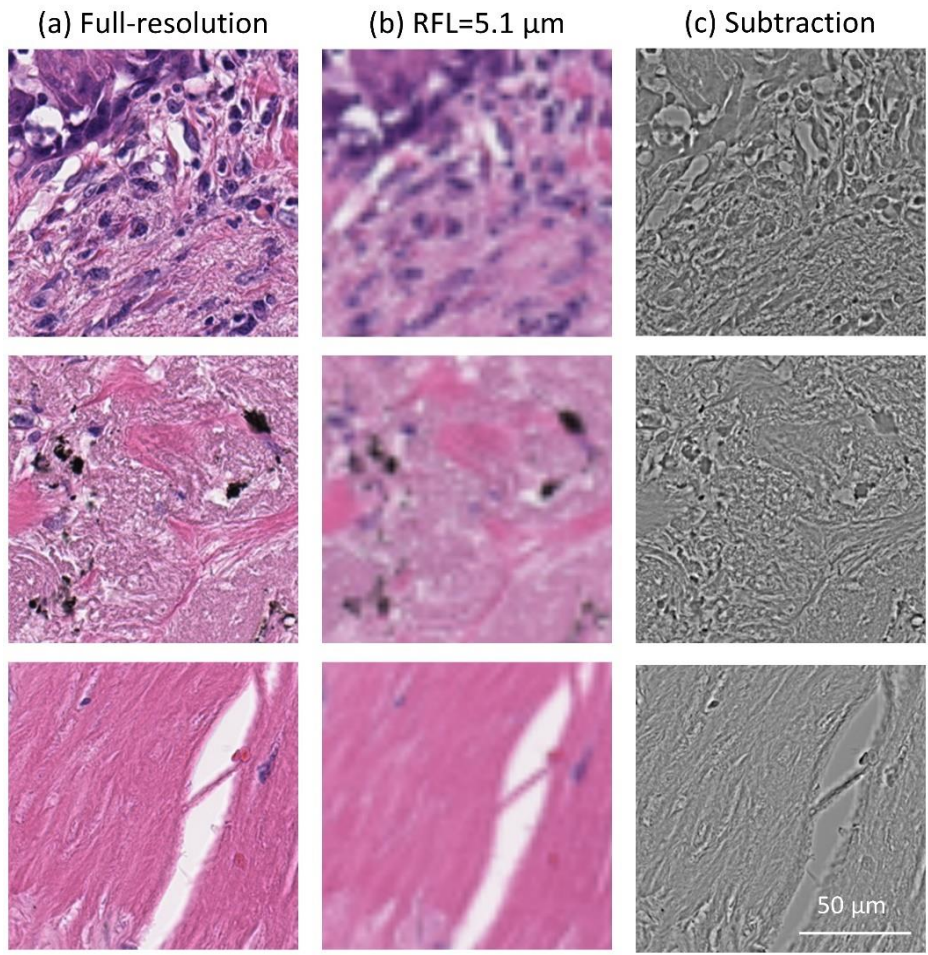

Fig. 3 Visualization of (a) original image patches, (b) images at RFL=5.1 µm, and (c) the subtraction of (a) and (b). Note that while color information is preserved, the architectural and cellular information is largely lost at RFL 5.1 microns, as demonstrated by the preservation of these features in the subtraction images.

Fig. 3 visually illustrates the features of interest. The left column shows representative vignettes of the original images at maximal resolution, that is RFL of 1.2 microns. The middle column shows the same vignettes with features of length-scale smaller than 5.1 microns removed. The differences between the two columns are shown in the right column. From Fig. 3, we can appreciate the type of visual data that DNN depends on for its prediction. Particularly notable here is that in while color information is preserved (which as mentioned earlier clearly carries information used by the DNN in its predictive algorithm), the architectural and cellular information is largely lost at RFL >5.1 microns.

Our MFL analysis shows a prediction accuracy transition as well. The curves for accuracy versus MFL underwent a transition at MFL of 41 microns – for ease of reference, we will label this as the characteristic MFL. This transition point was determined by minimizing the absolute residue error from the piecewise linear regression. This characteristic MFL indicates that the DNN is not simply focused on single cells (average size of 8 microns) for its prediction but instead appears to use stromal content and/or the inter-relationship between cells as well. As the MFL increases from 41 to 114 microns, there is a slight increase in information accuracy, approaching 80% at 114 microns. However, that increase is modest.

Taken together, these two sub-studies indicate that the DNN derives its predictive power from both sub-cellular content information and tissue-level content information. Neither set of information on their own is sufficient for DNN to obtain optimal predictive accuracy. These data suggests that, in the context of the current study parameters, the majority of information content relevant to DNN for prediction lies within RFL(resolution) of $\leq 1.2$ microns and MFL (field-of-view) $\geq 41$ microns). The data also strongly suggest that if we had been able to achieve an even higher RFL, that is <1.2 microns (which would be possible by scanning the slides at higher power), we could have achieved even higher DNN predictive accuracy.

This length-scale analysis can be used to further facilitate an understanding of DNN attention across the whole-slide images. To visualize the DNN attention across the whole slide image, we can produce slope maps through our RFL and MFL DNNs. For an RFL slope map (that is, a map indicating the magnitude of the positive or negative effect of that region on the DNN predictive accuracy), we start by focusing on a single image tile. We then adjust the image's RFL and feed the adjusted image through the appropriate trained DNN that is matched to the RFL. We repeat this process for the RFL from 1.2 microns to 5.1 microns (characteristic RFL). We then perform linear fitting and obtain a slope value. This slope value can be interpreted as a measure of the sensitivity of the DNN to the image features. The same process can then be repeated for all tiles in the whole slide image. The slopes of the linear fittings after normalization to [-1, +1] can be taken to represent the image tiles and used to generate an RFL sensitivity slope map with large values representing higher sensitivity. The generation of a corresponding MFL map follows the same approach. In this case, the process will use the MFL range from 2.5 to 41 microns (characteristic MFL).

We applied this slope analysis approach to two slides from our data set. The results are presented in Fig. 4 (Met+ case) and Fig. 5 (Met- case), both correctly predicted by the DNN. Figs. 4 and 5 (a-c) is a) the whole slide image, b) processed annotation mask and c) the corresponding primary tumor region. In the slope maps based on RFL (Figs. 4d and 5d) and MFL (Fig 4e and 5e) the warm (orange) color indicates tiles where the DNN prediction was accurate and sensitive to either RFL or MFL (Figs. 4d and 4e respectively), while cool (blue) color indicates tiles where the wrong prediction was made by that tile. Tiles with no color indicate areas where the DNN did not make a meaningful prediction and the prediction was insensitive to either RFL or MFL.

The blue colored tiles deserve some elaboration. For these tiles, the trained DNN actually made the wrong prediction on the tiles but made the correct prediction at the slide level. This dichotomy can be understood that as we begin to drop out image information (by increasing RFL or MFL), the trained DNN's prediction for a given tile began to asymptote towards a random guess prediction, which is on average closer to a correct prediction.

At the slide level, we can analyze the histologic features of the tiles demonstrating accurate prediction and sensitivity to changes in RFL and MFL (Figs. 4f and 5f) versus where the DNN did not make a meaningful prediction or an inaccurate prediction (Figs. 4g and 5g). For the tiles, the DNN was sensitive to changes in RFL and MFL and accurately predicted metastasis (Fig. 4f) or accurately predicted no metastasis (Fig. 5f), it can easily be seen that these tiles contain tumor and surrounding tumor microenvironment. In contrast, the histology seen in the tiles with no information for the DNN (i.e. where the DNN did not make a meaningful prediction or made an inaccurate prediction and was insensitive to changes in RFL and MFL, Figs. 4g and 5g) was very different; in these tiles, very little to no tumor or tumor microenvironment is seen, instead showing benign tissue composed of fibrosis, anthracitic pigment deposition, alveolar wall, fibroelastic tissue, pulmonary macrophages and/or reactive type pneumocytes.

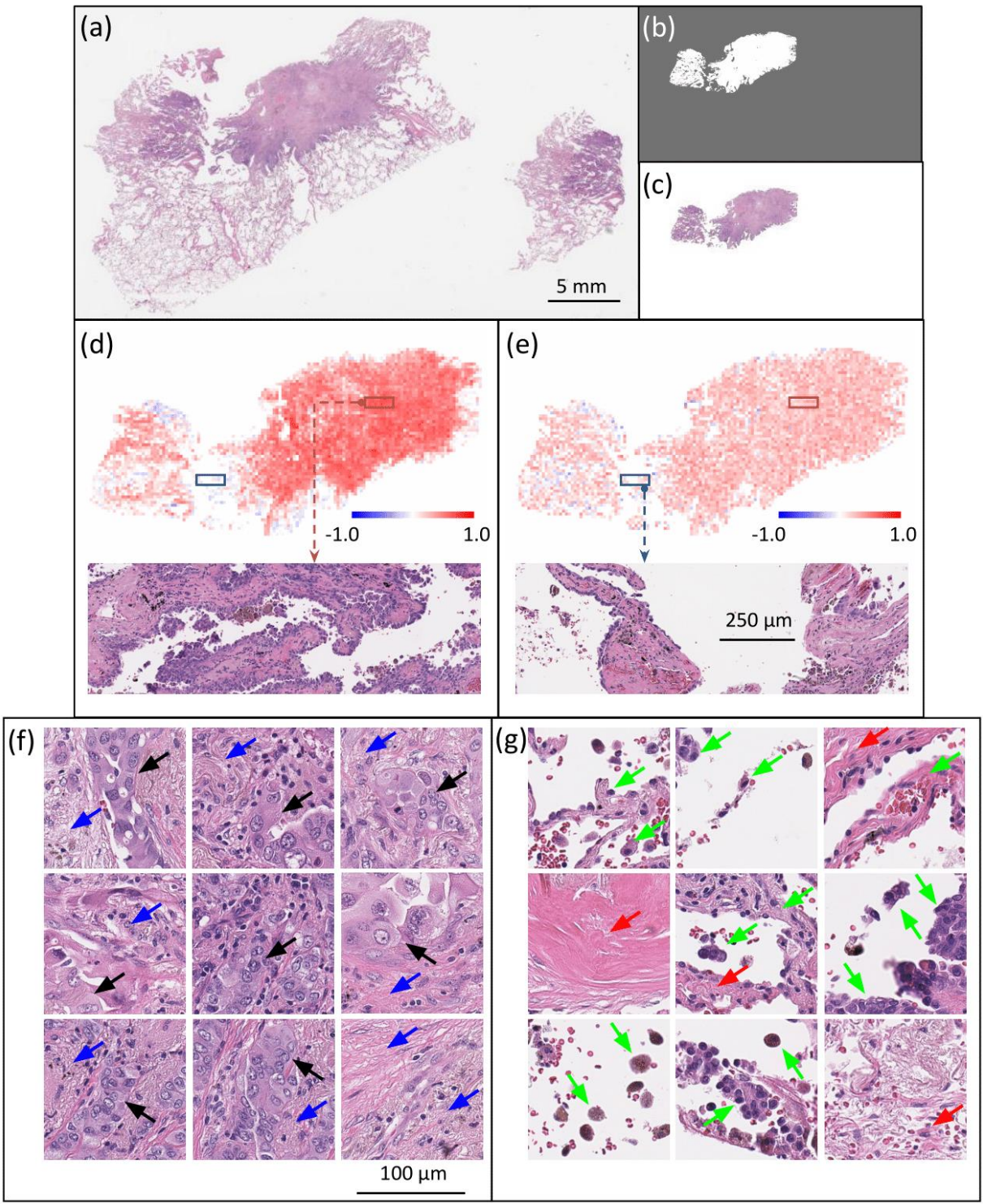

Fig. 4 (a) Whole slide image of a Met+ case. (b) Processed annotation mask. (c) Annotated H&E section. Slope maps for individual tiles examining the role of (d) RFLs and (e) MFLs, where orange indicates tiles where the DNN prediction for that tile was accurate (i.e. predicted brain metastasis) and sensitive to changes in RFL or MFL, blue indicating tiles where DNN made the incorrect prediction, and clear where the DNN was not able to make a meaningful prediction and where the tiles were insensitive to changes in RFL and MFL. In (d) and (e) the red and blue boxes show the same area of the tumor analyzed for RFL (d) and MFL (e); the red box shows a collection of tiles that demonstrated high predictive accuracy and sensitivity to RFL (d) and MFL (e), with the inset in (d) demonstrating the histology of that area. Note the presence of tumor and tumor microenvironment, further shown in other similar areas in (f). The blue box shows a collection of tiles that demonstrated no or negative predictive value and low sensitivity to RFL and MFL, with the inset in (e) demonstrating the histology of that area. Note that there are no tumor cells or tumor microenvironment, further shown in similar areas in (g). In (d) and (e) also note that the areas showing high predictive value and sensitivity to RFL and MFL largely overlap, and the areas showing low or negative predictive value and low sensitivity to RFL and MFL similarly overlap. Representative histologic images of tiles where (f) the DNN made an accurate prediction and was sensitive to changes in RFL and MFL and (g) the DNN made no prediction or an inaccurate prediction and was insensitive to changes in RFL and MFL. In (f), note areas of tumor (black arrows) and tumor microenvironment including immune cells and desmoplastic stroma (blue arrows). In (g) note that there are reactive pneumocytes, pulmonary macrophages and alveolar wall (green arrows) and fibrosis (red arrows) no tumor cells.

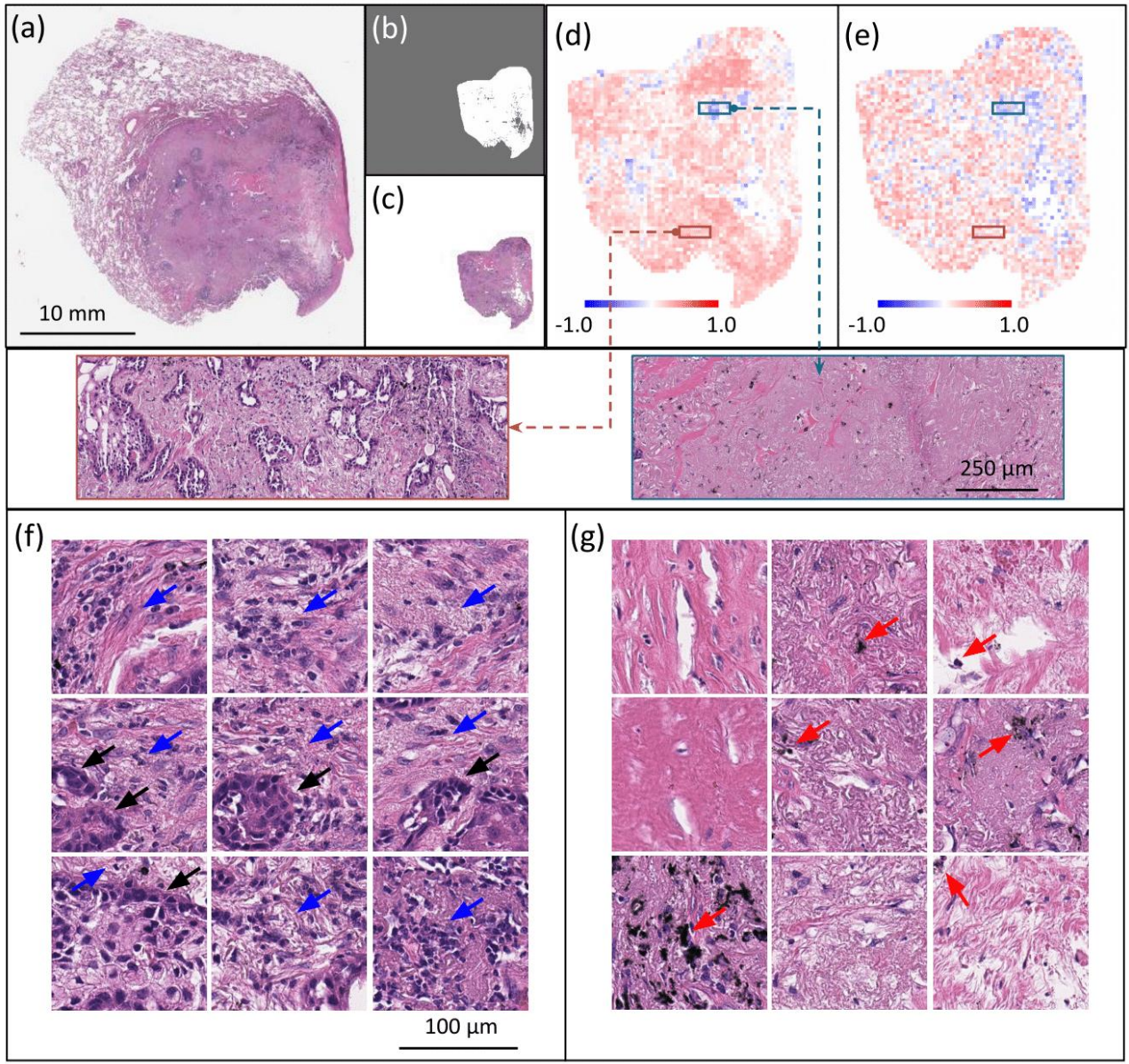

Fig. 5 (a) Whole slide image of a Met- case. (b) Processed annotation mask. (c) Annotated H&E section. Slope maps for individual tiles examining the role of (d) RFLs and (e) MFLs, where orange indicates tiles where the DNN prediction for that tile was accurate (i.e. predicted brain metastasis) and sensitive to changes in RFL or MFL, blue indicating tiles where DNN made the incorrect prediction, and clear where the DNN was not able to make a meaningful prediction and where the tiles were insensitive to changes in RFL and MFL. In (d) and (e) the red and blue boxes show the same area of the tumor analyzed for RFL (d) and MFL (e); the red box shows a collection of tiles that demonstrated high predictive accuracy and sensitivity to RFL (d) and MFL (e), with the inset in (d) demonstrating the histology of that area. Note the presence of tumor and tumor microenvironment, further shown in other similar areas in (f). The blue box shows a collection of tiles that demonstrated no or negative predictive value and low sensitivity to RFL and MFL, with the inset in (e) demonstrating the histology of that area. Note that there are no tumor cells or tumor microenvironment, further shown in similar areas in (g). In (d) and (e) also note that the areas showing high predictive value and sensitivity to RFL and MFL largely overlap, and the areas showing low or negative predictive value and low sensitivity to RFL and MFL similarly overlap. Representative histologic images of tiles where (f) the DNN made an accurate prediction and was sensitive to changes in RFL and MFL and (g) the DNN made no prediction or an inaccurate prediction and was insensitive to changes in RFL and MFL. In (f), note areas of tumor (black arrows) and tumor microenvironment including immune cells and desmoplastic stroma (blue arrows). In (g) note that there is fibrosis and deposits of anthracitic pigment (red arrows) but no tumor cells.

Beyond providing insights at the whole-slide level and image tile-level, the length-scale analysis method can also be applied to regional image patches, for example, patches comprised of concatenation of 5-by-5 image tiles. Using the slope maps, to take into account even wider tissue area than 41 microns we selected examples of regions with low and high RFL and MFL sensitivity (Figs. 6 and 7). These are the same slide cases as Figs. 4 and 5, respectively. In this analysis, we concatenated 5-by-5 adjacent image tiles on the cases demonstrated in Figs. 4 and 5 in areas showing high predictive accuracy and high RFL and MFL sensitivity vs areas showing low predictive accuracy and low sensitivity to RFL and MFL (Fig. 6, Met+ and Fig. 7, Met-). Figs. 6a and b and 7a and b demonstrate the histologic findings in Met+ and Met- cases where the individual tiles demonstrated accurate prediction and sensitivity to RFL (Figs. 6a2,b2 and Figs. 7a2,b2) and MFL (Figs. 6a3,b3 and Figs. 7a3,b3). The histology of these overall areas (Figs. 6a1,b1 and 7a1,b1) demonstrates clear tumor and peri-tumoral/tumor microenvironment, including desmoplastic stroma and infiltrating immune cells. In contrast, the areas where the concatenated tiles showed little prediction accuracy and insensitivity to RFL (Figs. 6c2,d2 and 7c2,d2) and MFL (Figs. 6c3,d3 and 7c3,d3) demonstrate an overall histology largely devoid of tumor cells and tumor stroma/microenvironment, consisting mainly of benign tissue including fibrosis, alveoli, reactive pneumocytes and macrophages (Figs. 6c1,d1 and 7c1,d1). Overall, these findings reinforce the idea that the DNN focuses on areas with tumors and that both increasing the RFL and decreasing the MFL negatively impact the ability of the DNN to correctly predict brain metastasis.

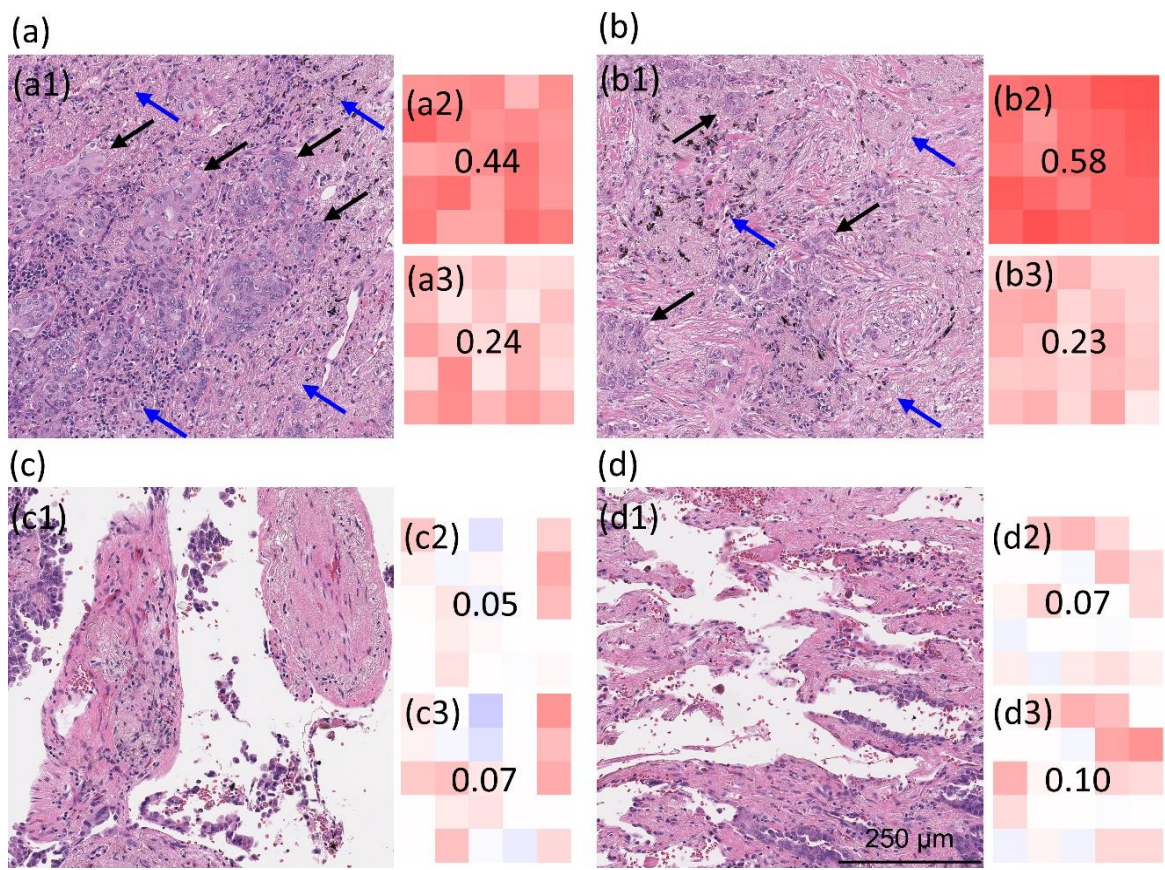

Fig. 6 (a-d) Concatenation of 5-by-5 image tiles from Met+ case in Figure 5. Areas a1 and b1 is the histology from a concatenated area where the DNN prediction was accurate and sensitive to changes in RFL (a2,b2) and MFL (a3,b3). Areas c1 and d1 is the histology from a concatenated area where the DNN prediction was low or inaccurate and insensitive to changes in RFL (c2,d2) and MFL (c3,d3). The color scheme is similar as for Figure 5. In a1 and b1, note areas of tumor cells (black arrows) and tumor microenvironment including immune cells and desmoplastic stroma (blue arrows). In c1 and d1, note that the tissue is largely devoid of tumor cells.

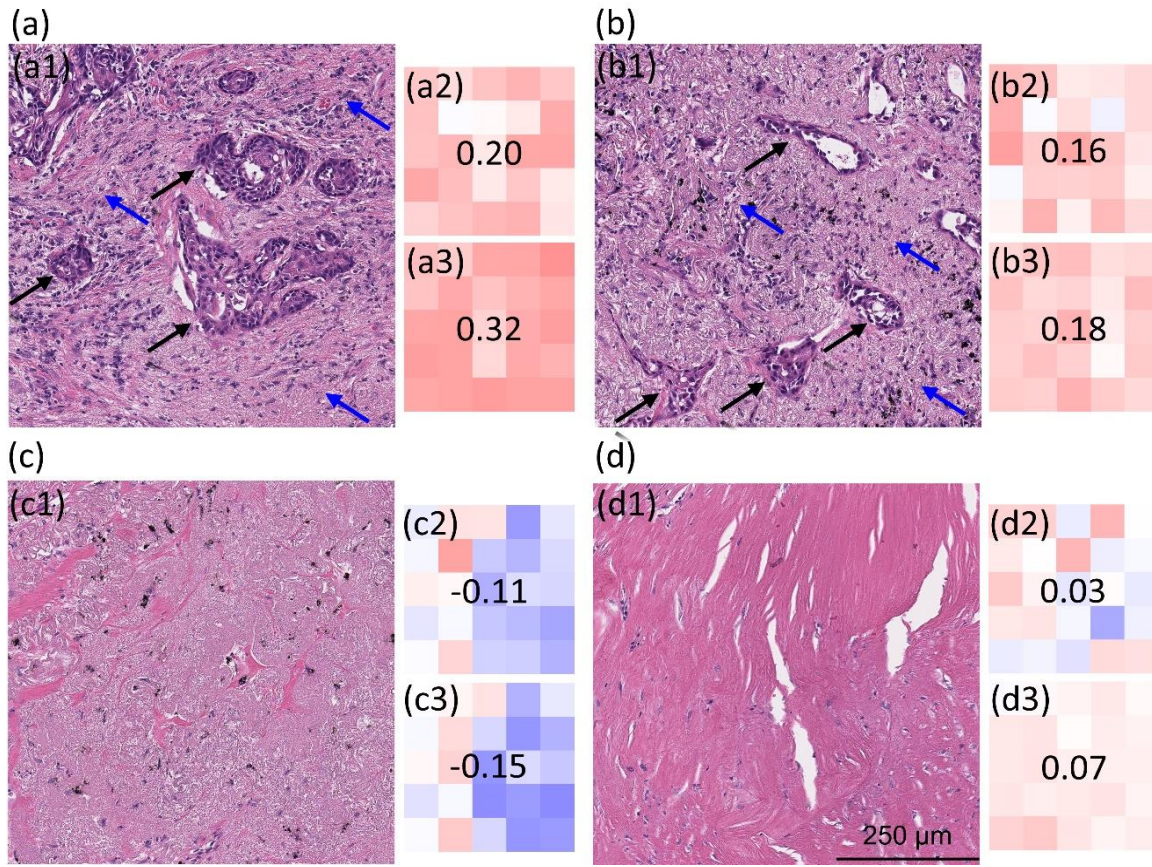

Fig. 7 (a-d) Concatenation of 5-by-5 image tiles from Met- case in Figure 6. Areas a1 and b1 is the histology from a concatenated area where the DNN prediction was accurate and sensitive to changes in RFL (a2,b2) and MFL (a3,b3). Areas c1 and d1 is the histology from a concatenated area where the DNN prediction was low or inaccurate and insensitive to changes in RFL (c2,d2) and MFL (c3,d3). The color scheme is similar as for Figure 6. In a1 and b1, note areas of tumor cells (black arrows) and tumor microenvironment including immune cells and desmoplastic stroma (blue arrows). In c1 and d1, note that the tissue is largely devoid of tumor cells.

## 4  Discussion

AI/DNN based learning is playing an increasingly important role in digital pathology. DNN have been trained to automatically and accurately identify known diagnostic histopathologic features that recapitulate the abilities of pathologists to identify these features [41], [42]. More recently, AI/DNN learning on digital images has provided exciting promise for predicting clinical outcomes by evaluation of routine H/E-stained sections in the prostate and several other tumor types [43]–[45]. The use of weakly-supervised learning to identify features that cannot be recognized by pathologists, such as progression and survival potential based on a routine histologic preparation, provides even more exciting and clinically impactful opportunities [43], [46]–[51]. Attention-based learning has also been utilized to analyze sub-regions of histopathology to identify patterns of highest diagnostic value. However, even with the use of attention maps and supervised learning, we do not understand what the AI/DNN is "learning" from, and there has been virtually no effort to understand the biology behind the AI/DNN learning process. Further, there has been little work in trying to understand the physical parameters required by AI/DNN to optimize its learning/predictive capability.

In this work, we present a method for assessing the parameters necessary to optimize DNN by analyzing the feature length-scale sensitivity of a trained image analysis DNN. Specifically, we examined the role of resolution (i.e. small sub-cellular scale features) via RFL and the role of more macro multi-cellular/tissue scale features via MFL. Using this approach, we examined a DNN algorithm that had been trained and validated to predict the future occurrence of brain metastases, or no metastases after long-term follow-up, in patients with early-stage (Stage I, II and III) NSCLC, based on DNN learning on the original H&E-stained slides of the diagnostic biopsies from these patients [29].

Several interesting features are observed. It is clear that the higher the resolution (that is, the smaller the RFL), the greater the accuracy of the DNN algorithm. In addition, we have also shown that a more macroscopic/multicellular tissue-based assessment (that is, larger MFL) is also crucial to DNN predictive capability. The observation that the higher resolving power (that is smaller RFL) than was possible in the current data set would result in even greater predictive accuracy strongly suggests that parameters could be optimized in future AI/DNN analysis of digital images, including capturing images at higher power magnification than was done in the current study (20X). In any event, our study clearly shows, for the first time to our knowledge, that it is a combination of sub-cellular and macro-cellular features that are important in DNN learning.

We have also observed, in this study and previously [29], that as we segment the DNN learning to small (tile) areas of defined pixel quantity, not all areas provide equal or even useful information. Indeed, in some cases, this tile based analysis shows areas where the DNN prediction based on that area is the opposite to the overall prediction, and the opposite to truth. In order to further assess this, we compared the histology of the areas that showed high sensitivity to the effects of RFL and MFL and provided predictive accuracy with areas that were insensitive to RFL and MFL and provided no or even negative predictive accuracy. Our analysis clearly demonstrates that the histologic features of these areas are quite distinct; the histology of the highly sensitive and predictive areas show tumor cells in association with tumor microenvironment, such as infiltrating immune (lymphoid) cells and desmoplastic response (which is a reaction that is highly specific for tumor growth [51]–[53]), and this is seen in the predictive areas in tumors where the DNN predicted no metastasis (and was correct) as well as in tumors where the DNN predicted the subsequent occurrence of brain metastases (and was correct). In contrast, the sub-areas in tumors that showed no or negative predictive value, but for which, at the slide level, the DNN correctly predicted the outcome for that patient (either the development of metastases or no metastases), these sub-areas were virtually devoid of tumor and elements of the tumor microenvironment, instead showing fibrosis, anthracosis, alveolar wall, pulmonary macrophages and reactive type pneumocytes. It is also notable that the areas showing high predictive value and sensitivity to RFL and MFL largely overlap and demonstrate tumor cells and tumor microenvironment, while the areas showing low or negative predictive value and low sensitivity to RFL and MFL similarly overlap and show no (or very few) tumor cells or tumor microenvironment. This clearly demonstrates that the presence of tumors is (not surprisingly) crucial in establishing a predictive DNN algorithm assessing tumor progression. What is perhaps more instructive is that tumor cells are not the only crucial feature, and elements of the tumor microenvironment appear to be independently important in establishing the predictive potential of the algorithm. We expect these observations to take on increasing importance as we dissect the biological basis for the ability of DNN to predict outcomes based on a digital image

In the broader context, for predictive tasks that are difficult/impossible for human experts to perform, but that deep learning/AI methods have demonstrated predictive value, we expect our method can help shed light on how different scales of features contribute to model predictions. As this method is agnostic to the DNN architecture, we anticipate that it can be applied broadly to gain insights in a whole host of DNN systems.

It is also likely that the characteristic RFL and MFL values will vary across disease types. For example, chronic diseases, such as diabetes, that have broader systemic impacts but that alter individual cell morphology minimally may have a low characteristic RFL value and a high MFL value. Finally, this study should contribute to the rational design of future studies to apply deep learning/AI to digital histopathologic images, including guidance on techniques of image acquisition and field of view metrics.

**Acknowledgement**

This study was supported by U01CA233363 from the National Cancer Institute (RJC) and by the Washington University in St. Louis School of Medicine Personalized Medicine Initiative (RJC). HZ, SL, SM and CY are also supported by Sensing to Intelligence (S2I) (Grant No. 13520296) and Heritage Research Institute for the Advancement of Medicine and Science at Caltech (Grant No. HMRI-15-09-01). MW and RG are also supported 5R01CA182746 from the National Cancer Institute.

**Conflict of Interest**

The authors declare no conflict of interest.

**References**

[1] A. Janowczyk and A. Madabhushi, "Deep learning for digital pathology image analysis: A comprehensive tutorial with selected use cases," Journal of Pathology Informatics, vol. 7, no. 1, p. 29, Jan. 2016. [Online]. Available: https://linkinghub.elsevier.com/retrieve/ pii/S2153353922005478

[2] V. Baxi, R. Edwards, M. Montalto, and S. Saha, "Digital pathology and artificial intelligence in translational medicine and clinical practice," Modern Pathology, vol. 35, no. 1, pp. 23–32, Jan. 2022. [Online]. Available: https://linkinghub.elsevier.com/retrieve/pii/ S0893395222003477

[3] C. Shen, S. Rawal, R. Brown, H. Zhou, A. Agarwal, M. A. Watson, R. J. Cote, and C. Yang, "Automatic detection of circulating tumor cells and cancer associated fibroblasts using deep learning," Scientific Reports, vol. 13, no. 1, p. 5708, Apr. 2023. [Online]. Available: https://www.nature.com/articles/s41598-023-32955-0

[4] T. Falk, D. Mai, R. Bensch, O. Cicek, A. Abdulkadir, Y. Marrakchi, A. Bohm, J. Deubner, Z. J¨ackel, K. Seiwald, A. Dovzhenko, O. Tietz, ¨C. Dal Bosco, S. Walsh, D. Saltukoglu, T. L. Tay, M. Prinz, K. Palme, M. Simons, I. Diester, T. Brox, and O. Ronneberger, "U-Net: deep learning for cell counting, detection, and morphometry," Nature Methods, vol. 16, no. 1, pp. 67–70, Jan. 2019. [Online]. Available: https://www.nature.com/articles/s41592-018-0261-2

[5] E. Mohamed, K. Sirlantzis, and G. Howells, "A review of visualisationas-explanation techniques for convolutional neural networks and their evaluation," Displays, vol. 73, p. 102239, Jul. 2022. [Online]. Available: https://linkinghub.elsevier.com/retrieve/pii/S014193822200066X

[6] Q. Teng, Z. Liu, Y. Song, K. Han, and Y. Lu, "A survey on the interpretability of deep learning in medical diagnosis," Multimedia Systems, vol. 28, no. 6, pp. 2335–2355, Dec. 2022. [Online]. Available: https://link.springer.com/10.1007/s00530-022-00960-4

[7] A. Singh, S. Sengupta, and V. Lakshminarayanan, "Explainable Deep Learning Models in Medical Image Analysis," Journal of Imaging, vol. 6, no. 6, p. 52, Jun. 2020. [Online]. Available: https://www.mdpi.com/2313-433X/6/6/52

[8] Z. Salahuddin, H. C. Woodruff, A. Chatterjee, and P. Lambin, "Transparency of deep neural networks for medical image analysis: A review of interpretability methods," Computers in Biology and Medicine, vol. 140, p. 105111, Jan. 2022. [Online]. Available: https://linkinghub.elsevier.com/retrieve/pii/S0010482521009057

[9] K. Simonyan, A. Vedaldi, and A. Zisserman, "Deep Inside Convolutional Networks: Visualising Image Classification Models and Saliency Maps," arXiv, 2013, publisher: arXiv Version Number: 2. [Online]. Available: https://arxiv.org/abs/1312.6034

[10] R. R. Selvaraju, M. Cogswell, A. Das, R. Vedantam, D. Parikh, and D. Batra, "Grad-CAM: Visual Explanations from Deep Networks via Gradient-Based Localization," in 2017 IEEE International Conference on Computer Vision (ICCV). Venice: IEEE, Oct. 2017, pp. 618–626. [Online]. Available: http://ieeexplore.ieee.org/document/8237336/

[11] J. T. Springenberg, A. Dosovitskiy, T. Brox, and M. Riedmiller, "Striving for Simplicity: The All Convolutional Net," arXiv, 2014, publisher: [object Object] Version Number: 3. [Online]. Available: https://arxiv.org/abs/1412.6806

[12] H. Chen, S. M. Lundberg, and S.-I. Lee, "Explaining a series of models by propagating Shapley values," Nature Communications, vol. 13, no. 1, p. 4512, Aug. 2022. [Online]. Available: https://www.nature.com/articles/s41467-022-31384-3

[13] M. D. Zeiler and R. Fergus, "Visualizing and Understanding Convolutional Networks," in Computer Vision – ECCV 2014, D. Fleet, T. Pajdla, B. Schiele, and T. Tuytelaars, Eds. Cham: Springer International Publishing, 2014, vol. 8689, pp. 818–833, series Title: Lecture Notes in Computer Science. [Online]. Available: http://link.springer.com/10.1007/978-3-319-10590-1_53

[14] I. J. Goodfellow, J. Shlens, and C. Szegedy, "Explaining and Harnessing Adversarial Examples," arXiv, 2014, publisher: arXiv Version Number: 3. [Online]. Available: https://arxiv.org/abs/1412.6572

[15] Q. Zhang, Y. N. Wu, and S.-C. Zhu, "Interpretable Convolutional Neural Networks," in 2018 IEEE/CVF Conference on Computer Vision and Pattern Recognition. Salt Lake City, UT: IEEE, Jun. 2018, pp. 8827–8836. [Online]. Available: https://ieeexplore.ieee.org/document/8579018/

[16] J. Yosinski, J. Clune, A. Nguyen, T. Fuchs, and H. Lipson, "Understanding Neural Networks Through Deep Visualization," arXiv, 2015, publisher: arXiv Version Number: 1. [Online]. Available: https://arxiv.org/abs/1506.06579

[17] B. Zhou, D. Bau, A. Oliva, and A. Torralba, "Interpreting Deep Visual Representations via Network Dissection," IEEE Transactions on Pattern Analysis and Machine Intelligence, vol. 41, no. 9, pp. 2131–2145, Sep. 2019. [Online]. Available: https://ieeexplore.ieee.org/document/8417924/

[18] P. W. Koh, T. Nguyen, Y. S. Tang, S. Mussmann, E. Pierson, B. Kim, and P. Liang, "Concept bottleneck models," in International conference on machine learning. PMLR, 2020, pp. 5338–5348.

[19] Y. Dai, G. Wang, and K.-C. Li, "Conceptual alignment deep neural networks," Journal of Intelligent & Fuzzy Systems, vol. 34, no. 3, pp. 1631–1642, Mar. 2018. [Online]. Available: https://www.medra.org/servlet/aliasResolver?alias=iospress&doi=10.3233/JIFS-169457

[20] S. Shen, S. X. Han, D. R. Aberle, A. A. Bui, and W. Hsu, "An interpretable deep hierarchical semantic convolutional neural network for lung nodule malignancy classification," Expert Systems with Applications, vol. 128, pp. 84–95, Aug. 2019. [Online]. Available: https://linkinghub.elsevier.com/retrieve/pii/S0957417419300545

[21] O. Li, H. Liu, C. Chen, and C. Rudin, "Deep Learning for Case-Based Reasoning Through Prototypes: A Neural Network That Explains Its Predictions," Proceedings of the AAAI Conference on Artificial Intelligence, vol. 32, no. 1, Apr. 2018. [Online]. Available: https://ojs.aaai.org/index.php/AAAI/article/view/11771

[22] C. Chen, O. Li, D. Tao, A. Barnett, C. Rudin, and J. K. Su, "This Looks Like That: Deep Learning for Interpretable Image Recognition," in Advances in Neural Information Processing Systems, H. Wallach, H. Larochelle, A. Beygelzimer, F. d. Alche-Buc, ´E. Fox, and R. Garnett, Eds., vol. 32. Curran Associates, Inc., 2019. [Online]. Available: https://proceedings.neurips.cc/paper files/ paper/2019/file/adf7ee2dcf142b0e11888e72b43fcb75-Paper.pdf

[23] Q. H. Cao, T. T. H. Nguyen, V. T. K. Nguyen, and X. P. Nguyen, "A Novel Explainable Artificial Intelligence Model in Image Classification problem," arXiv, 2023, publisher: arXiv Version Number: 1. [Online]. Available: https://arxiv.org/abs/2307.04137

[24] B. H. Van Der Velden, H. J. Kuijf, K. G. Gilhuijs, and M. A. Viergever, "Explainable artificial intelligence (XAI) in deep learningbased medical image analysis," Medical Image Analysis, vol. 79, p. 102470, Jul. 2022. [Online]. Available: https://linkinghub.elsevier.com/retrieve/pii/S1361841522001177

[25] S. Atakishiyev, M. Salameh, H. Yao, and R. Goebel, "Explainable Artificial Intelligence for Autonomous Driving: A Comprehensive Overview and Field Guide for Future Research Directions," Apr. 2024, arXiv:2112.11561 [cs]. [Online]. Available: http://arxiv.org/abs/2112. 11561

[26] A. K. Ganti, A. B. Klein, I. Cotarla, B. Seal, and E. Chou, "Update of Incidence, Prevalence, Survival, and Initial Treatment in Patients With Non–Small Cell Lung Cancer in the US," JAMA Oncology, vol. 7, no. 12, p. 1824, Dec. 2021. [Online]. Available: https://jamanetwork.com/journals/jamaoncology/fullarticle/2784988

[27] S. N. Waqar, D. Morgensztern, and R. Govindan, "Systemic Treatment of Brain Metastases," Hematology/Oncology Clinics of North America, vol. 31, no. 1, pp. 157–176, Feb. 2017. [Online]. Available: https://linkinghub.elsevier.com/retrieve/pii/S0889858816301216

[28] D. C. C. Tsui, D. R. Camidge, and C. G. Rusthoven, "Managing Central Nervous System Spread of Lung Cancer: The State of the Art," Journal of Clinical Oncology, vol. 40, no. 6, pp. 642–660, Feb. 2022. [Online]. Available: https://ascopubs.org/doi/10.1200/JCO.21.01715

[29] H. Zhou, M. Watson, C. T. Bernadt, S. S. Lin, C. Lin, J. H. Ritter, A. Wein, S. Mahler, S. Rawal, R. Govindan, C. Yang, and R. J. Cote, "Ai-guided histopathology predicts brain metastasis in lung cancer patients," The Journal of Pathology, vol. 263, no. 1, pp. 89–98, May 2024. [Online]. Available: https://pathsocjournals.onlinelibrary. wiley.com/doi/10.1002/path.6263

[30] N. Otsu, "A Threshold Selection Method from Gray-Level Histograms," IEEE Transactions on Systems, Man, and Cybernetics, vol. 9, no. 1, pp. 62–66, Jan. 1979. [Online]. Available: http://ieeexplore.ieee.org/ document/4310076/

[31] A. Vahadane, T. Peng, A. Sethi, S. Albarqouni, L. Wang, M. Baust, K. Steiger, A. M. Schlitter, I. Esposito, and N. Navab, "StructurePreserving Color Normalization and Sparse Stain Separation for Histological Images," IEEE Transactions on Medical Imaging, vol. 35, no. 8, pp. 1962–1971, Aug. 2016. [Online]. Available: http://ieeexplore.ieee.org/document/7460968/

[32] K. He, X. Zhang, S. Ren, and J. Sun, "Deep Residual Learning for Image Recognition," arXiv, 2015, publisher: arXiv Version Number: 1. [Online]. Available: https://arxiv.org/abs/1512.03385

[33] J. M. Johnson and T. M. Khoshgoftaar, "Survey on deep learning with class imbalance," Journal of Big Data, vol. 6, no. 1, p. 27, Mar. 2019. [Online]. Available: https://doi.org/10.1186/s40537-019-0192-5

[34] S. H. Khan, M. Hayat, M. Bennamoun, F. A. Sohel, and R. Togneri, "Cost-sensitive learning of deep feature representations from imbalanced data," IEEE Transactions on Neural Networks and Learning Systems, vol. 29, no. 8, pp. 3573–3587, 2018.

[35] H. Wang, Z. Cui, Y. Chen, M. Avidan, A. B. Abdallah, and A. Kronzer, "Predicting hospital readmission via cost-sensitive deep learning," IEEE/ACM Transactions on Computational Biology and Bioinformatics, vol. 15, no. 6, pp. 1968–1978, 2018.

[36] K. Nemoto, R. Hamaguchi, T. Imaizumi, and S. Hikosaka, "Classification of rare building change using cnn with multi-class focal loss," in IGARSS 2018 - 2018 IEEE International Geoscience and Remote Sensing Symposium, 2018, pp. 4663–4666.

[37] C. Zhang, K. C. Tan, and R. Ren, "Training cost-sensitive deep belief networks on imbalance data problems," in 2016 International Joint Conference on Neural Networks (IJCNN), 2016, pp. 4362–4367.

[38] Y. Zhang, L. Shuai, Y. Ren, and H. Chen, "Image classification with category centers in class imbalance situation," in 2018 33rd Youth Academic Annual Conference of Chinese Association of Automation (YAC), 2018, pp. 359–363.

[39] E. T. Whittaker, "XVIII.—On the Functions which are represented by the Expansions of the Interpolation-Theory," Proceedings of the Royal Society of Edinburgh, vol. 35, pp. 181–194, 1915. [Online]. Available: https://www.cambridge.org/core/product/identifier/ S0370164600017806/type/journal article

[40] G. Visona, L. M. Spiller, S. Hahn, E. Hattingen, T. J. Vogl, ` G. Schweikert, K. Bankov, M. Demes, H. Reis, P. Wild, P. S. Zeiner, F. Acker, M. Sebastian, and K. J. Wenger, "MachineLearning-Aided Prediction of Brain Metastases Development in Non–Small-Cell Lung Cancers," Clinical Lung

Cancer, vol. 24, no. 8, pp. e311–e322, Dec. 2023. [Online]. Available: https://linkinghub.elsevier.com/retrieve/pii/S1525730423001481

[41] G. Campanella, M. G. Hanna, L. Geneslaw, A. Miraflor, V. Werneck Krauss Silva, K. J. Busam, E. Brogi, V. E. Reuter, D. S. Klimstra, and T. J. Fuchs, "Clinical-grade computational pathology using weakly supervised deep learning on whole slide images," Nature Medicine, vol. 25, no. 8, pp. 1301–1309, Aug. 2019. [Online]. Available: https://www.nature.com/articles/s41591-019-0508-1

[42] M. F. A. Fauzi, W. Chen, D. Knight, H. Hampel, W. L. Frankel, and M. N. Gurcan, "Tumor Budding Detection System in Whole Slide Pathology Images," Journal of Medical Systems, vol. 44, no. 2, p. 38, Feb. 2020. [Online]. Available: http://link.springer.com/10.1007/s10916-019-1515-y

[43] P.-C. Tsai, T.-H. Lee, K.-C. Kuo, F.-Y. Su, T.-L. M. Lee, E. Marostica, T. Ugai, M. Zhao, M. C. Lau, J. P. Vayrynen, ¨ M. Giannakis, Y. Takashima, S. M. Kahaki, K. Wu, M. Song, J. A. Meyerhardt, A. T. Chan, J.-H. Chiang, J. Nowak, S. Ogino, and K.-H. Yu, "Histopathology images predict multi-omics aberrations and prognoses in colorectal cancer patients," Nature Communications, vol. 14, no. 1, p. 2102, Apr. 2023. [Online]. Available: https://www.nature.com/articles/s41467-023-37179-4

[44] J. Liang, W. Zhang, J. Yang, M. Wu, Q. Dai, H. Yin, Y. Xiao, and L. Kong, "Deep learning supported discovery of biomarkers for clinical prognosis of liver cancer," Nature Machine Intelligence, vol. 5, no. 4, pp. 408–420, Apr. 2023. [Online]. Available: https://www.nature.com/articles/s42256-023-00635-3

[45] A. Echle, N. T. Rindtorff, T. J. Brinker, T. Luedde, A. T. Pearson, and J. N. Kather, "Deep learning in cancer pathology: a new generation of clinical biomarkers," British Journal of Cancer, vol. 124, no. 4, pp. 686–696, Feb. 2021. [Online]. Available: https://www.nature.com/articles/s41416-020-01122-x

[46] N. Coudray, P. S. Ocampo, T. Sakellaropoulos, N. Narula, M. Snuderl, D. Fenyo, A. L. Moreira, N. Razavian, and A. Tsirigos, "Classification ¨ and mutation prediction from non–small cell lung cancer histopathology images using deep learning," Nature Medicine, vol. 24, no. 10, pp. 1559–1567, Oct. 2018. [Online]. Available: https://www.nature.com/articles/s41591-018-0177-5

[47] Z. Su, M. K. K. Niazi, T. E. Tavolara, S. Niu, G. H. Tozbikian, R. Wesolowski, and M. N. Gurcan, "BCR-Net: A deep learning framework to predict breast cancer recurrence from histopathology images," PLOS ONE, vol. 18, no. 4, p. e0283562, Apr. 2023. [Online]. Available: https://dx.plos.org/10.1371/journal.pone.0283562

[48] A. Esteva, J. Feng, D. Van Der Wal, S.-C. Huang, J. P. Simko, S. DeVries, E. Chen, E. M. Schaeffer, T. M. Morgan, Y. Sun, A. Ghorbani, N. Naik, D. Nathawani, R. Socher, J. M. Michalski, M. Roach, T. M. Pisansky, J. M. Monson, F. Naz, J. Wallace, M. J. Ferguson, J.-P. Bahary, J. Zou, M. Lungren, S. Yeung, A. E. Ross, NRG Prostate Cancer AI Consortium, M. Kucharczyk, L. Souhami, L.

Ballas, C. A. Peters, S. Liu, A. G. Balogh, P. D. Randolph-Jackson, D. L. Schwartz, M. R. Girvigian, N. G. Saito, A. Raben, R. A. Rabinovitch, K. Katato, H. M. Sandler, P. T. Tran, D. E. Spratt, S. Pugh, F. Y. Feng, and O. Mohamad, "Prostate cancer therapy personalization via multi-modal deep learning on randomized phase III clinical trials," npj Digital Medicine, vol. 5, no. 1, p. 71, Jun. 2022. [Online]. Available: https://www.nature.com/articles/s41746-022-00613-w

[49] K.-H. Yu, C. Zhang, G. J. Berry, R. B. Altman, C. Re, D. L. Rubin, ´ and M. Snyder, "Predicting non-small cell lung cancer prognosis by fully automated microscopic pathology image features," Nature Communications, vol. 7, no. 1, p. 12474, Aug. 2016. [Online]. Available: https://www.nature.com/articles/ncomms12474

[50] D. Bychkov, N. Linder, R. Turkki, S. Nordling, P. E. Kovanen, C. Verrill, M. Walliander, M. Lundin, C. Haglund, and J. Lundin, "Deep learning based tissue analysis predicts outcome in colorectal cancer," Scientific Reports, vol. 8, no. 1, p. 3395, Feb. 2018. [Online]. Available: https://www.nature.com/articles/s41598-018-21758-3

[51] G. M. Ratnayake, F.-M. Laskaratos, D. Mandair, M. E. Caplin, K. Rombouts, and C. Toumpanakis, "What Causes Desmoplastic Reaction in Small Intestinal Neuroendocrine Neoplasms?" Current Oncology Reports, vol. 24, no. 10, pp. 1281–1286, Oct. 2022. [Online]. Available: https://link.springer.com/10.1007/s11912-022-01211-5

[52] R. A. Walker, "The complexities of breast cancer desmoplasia," Breast Cancer Research, vol. 3, no. 3, p. 143, Jun. 2001. [Online]. Available: http://breast-cancer-research.biomedcentral.com/articles/10.1186/bcr287

[53] A. C. Martins Cavaco, S. Damaso, S. Casimiro, and L. Costa, ˆ "Collagen biology making inroads into prognosis and treatment of cancer progression and metastasis," Cancer and Metastasis Reviews, vol. 39, no. 3, pp. 603–623, Sep. 2020. [Online]. Available: https://link.springer.com/10.1007/s10555-020-09888-5